# Peculiar long-term fluorescence of Rb atoms in coated vapor cell with internal atomic source

S. N. Atutov[*], V. A. Sorokin

Institute of Automation and Electrometry SB RAS,
Koptyug Ave. 1, 630090 Novosibirsk, Russia
* atutov@fe.infn.it

S. N. Bagayev, M.N.Skvortsov, A.V. Taichenachev

Institute of Laser Physics SB RAS
Ac. Lavrentieva Ave.15B, 630090 Novosibirsk, Russia

We report on an experiment in which the fluorescence decay time of 5P levels of Rb atoms in a coated vapor cell exceeds several of milliseconds, many orders of magnitude longer than normal decay time of excited of rubidium atoms. We found that this peculiarly long - term decay is not exponential and can only be observed in a high quality coated cell with a long lifetime of optical pumped atoms on atomic ground-state sublevels. A numerical simulation based on a complete density matrix is presented. It supports the population pumping process through the ground-state sublevels of Rb atoms as the most plausible mechanism for the observed phenomenon and it consistently demonstrates all the relevant features of the experiment. The investigation can be used, for example, to explore the quality of a various anti-relaxation coatings used for the resonant cells.

Since its discovery in the 1950s optical pumping has been widely studied and used in atomic physics for a variety of applications. For example, optical pumping has been used for the coherent trapping of atomic populations [1], electromagnetically induced transparency and electromagnetically induced absorption, [2-4], Moreover, the optical pumping effect has been used for the storage of light [5], atomic clocks [6] and magnetometers [7], for the production of polarized ions [8–10], neutrons [11], solids [12] and for gas targets in nuclear physics studies [13], and for the generation of polarized noble gases through spin-exchange optical pumping [14].

In spite of the wide variety of activity in the field, the properties of coherently driven atomic media have not yet been studied completely. Most investigations are based on the use of spectrally narrow, low-power lasers and a coated or non-coated resonant glass cell filled with alkali metal vapor and with little or no buffer gas. It is well known that when two CW lasers are tuned so that they couple two different ground-state sublevels into a common upper level, there is no steady-state population in the upper level. This effect, which holds good even for intense laser fields, is termed coherent trapping of atomic populations. If, in this case, a frequency of one of the lasers is adiabatically shifted far away from the corresponding optical transition, fluorescence appears shortly thereafter due to the fast growing population on the common upper level (as a result of the destruction of the coherent trapping of the atomic population ground-state sublevels) after which it decays in accordance with the particular life-time of the excited atoms. Here we show that under similar conditions using two moderately power lasers, a large fluorescence feature with a decay time exceeding several milliseconds is achieved. The observed decay time is much longer than the characteristic time of the normal decay of the excited atoms. In our study, the behavior of the fluorescent spectrum is investigated to extract a wide range of the probe laser intensity and the detuning of the pump laser. A numerical simulation based on a complete density matrix is presented. It consistently demonstrates all the relevant features of the experiment.

The experimental setup is shown in Fig. 1. The populations of Rb vapor on 5S ground levels and the dynamics of the optical pumping process in the cell (1) are measured by fluorescence excited atoms using a probing free-running laser (2). In the optical pumping experiments, a separate pumping free-running passively-stabilized laser (3) is used. It has an expanded beam diameter of 10 mm oriented perpendicularly to the probing beam. Both lasers have moderate and equal powers of output beams (30 mW). Fluorescence intensity is measured by a photodiode (FD) which is connected to a data acquisition system (DAQ).

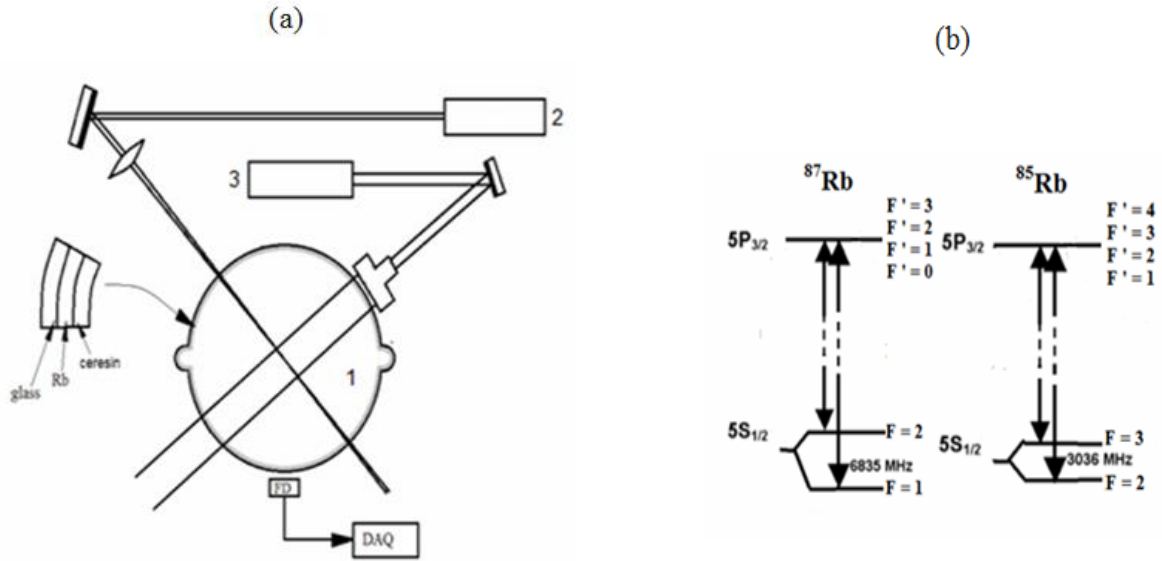

Figure 1 (a) experimental set up, 1 – cell, 2 - probing diode laser, 3 – pumping diode laser, FD – photodiode, DAQ -data acquisition system; (b) energy level diagram of $^{87}$Rb and $^{85}$Rb isotope.

To obtain the fluorescence spectra of the vapor, we periodically scan the probing laser frequency across all spectral lines of both isotopes of the Rb atoms using a modulation in the laser current. To reduce the influence of the probing radiation on the forms of spectral lines and the dynamics of optical pumping, the probing laser beam is focused strictly inside the cell (diameter of 100 m$\mu$). Because of the beam focusing and the frequency scanning, the Rb atoms interact with the probing laser radiation for a short time, too short in fact to pump the atoms effectively. In certain experiments, the powers of both lasers have been attenuated from their maximum power down to a few mW using a set of optical filters. In the experiment we have used a spherical cell (radius R = 6 cm) with an internal atomic source and coated with ceresin. Ceresin is a mixture of saturated hydrocarbons with a number of carbon atoms in the molecule varying from 36 to 55. It has a molecular mass of about 700 and melting temperature of ~ 120 $^0$C. The cell is an evacuated glass bulb characterized by the fact the source of atomic vapors in the form of a metal film of Rb has been evenly distributed throughout the inner surface of the bulb, and the ceresin film has also been distributed uniformly over this entire area and over the metal surface. In contrast to the cells in a conventional configuration, our cell has no separate atomic source consisting of appendix with piece of alkaline metal inside that would cause extra loss of optically pumped atoms due to their leaking from the bulb into the atomic source. This has allowed us to minimize the total loss of the optically pumped atoms in the cell. Minimal loss combined with the large diameter of the cell has provided the pumped atoms on the atomic ground-state sublevels with a long lifetime. The construction of the cell, coating procedure and properties of the cell have been described in a recent publication [15]. During all experiments the cell has been kept at room temperature, corresponding to a vapor pressure of ~ 6 × 10$^9$ cm$^{-3}$ that has assured an optically thin medium in the cell [16]. In certain experiments a paraffin-coated cell of conventional configuration has been used.

At the beginning of the experiment, we blocked the pumping laser beam and recorded the fluorescence in the cell excited by the probing laser. The fluorescence recorded as a function of time is presented in Figure 2. This Figure demonstrates the fluorescence spectrum of Rb atoms that is commonly observed in many experiments. The frequency of the probe laser was scanned over four optical transitions of both isotopes of Rb atoms several times; the single 12 GHz scan is indicated in the figure by the two vertical turning lines. The two natural isotopes of Rb, $^{85}$Rb and $^{87}$Rb, have ground - state hyperfine splittings of ≈3.0 and ≈6.8 GHz, respectively, that gave rise to four spectral peaks. The difference in the heights of the corresponding spectral peaks near the two turning lines was caused by residual optical pumping by the probing laser radiation, the frequency of which was scanned forward and backward.

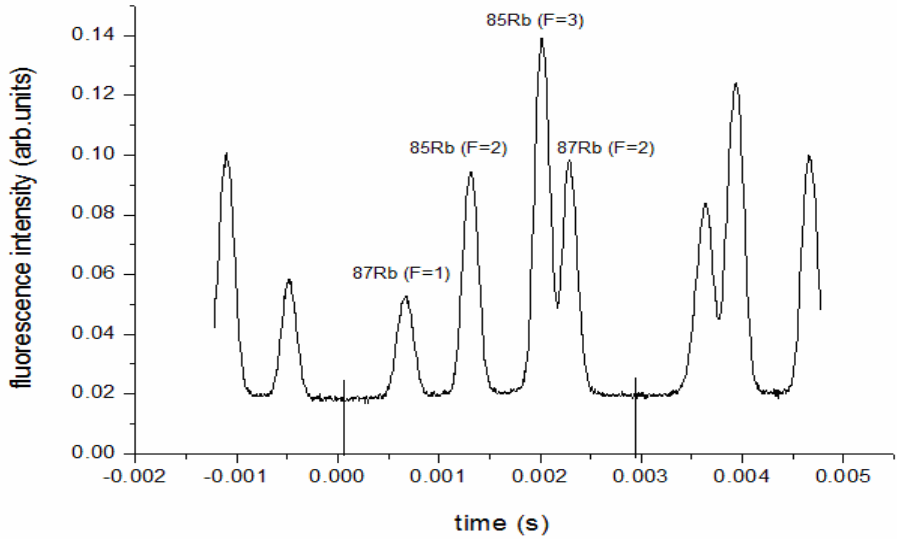

Figure 2. Fluorescence spectrum recorded as a function of time; two vertical lines indicate the single scan.

The frequency of the pumping laser was tuned to be resonant to $5S_{1/2} (F = 3) \rightarrow 5P_{3/2}$ of $^{85}$Rb optical transition, until a maximum of fluorescent intensity in the separate cell was obtained and then the pumping laser frequency was passively stabilized at this spectral point (see, Figure 1, b). The pump beam in the bulb was abruptly opened by a shutter and the resulting spectrum was as a function of time. This spectrum is presented in Figure 3. In the picture, the spectral position of the pump frequency is indicated by a solid line and the single scan of the probing laser is again indicated by two vertical lines. The atoms localized on the $^{87}$Rb atom ground-state sublevel $5S_{1/2} (F = 3)$ were excited by the pumping radiation and this populated $5S_{1/2} (F = 2)$ ground-state sublevel through the intermediate atomic upper states $5P_{3/2}$. Figure 3 shows that, as a result of the optical pumping process, the height of the $^{85}$Rb ($F = 3$) peak decreased, while the height of $^{85}$Rb ($F = 2$) peak increased. But, in addition to these deformed and relatively narrow spectral peaks of the $^{85}$Rb atom, Figure 3 shows a surprisingly large feature that according to the picture decayed within the time span of a few milliseconds. We also see slightly deformed peaks of the $^{87}$Rb isotope that do not make a noticeable contribution to the observed feature.

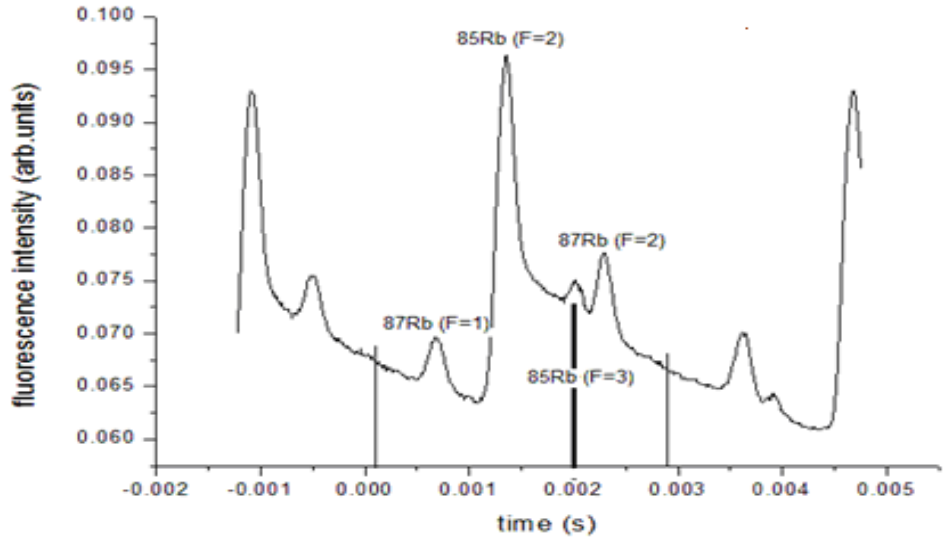

Figure 3. Fluorescence spectrum recorded as a function of time when $5S_{1/2}\,(F = 3) \rightarrow 5P_{3/2}$ optical transition of $^{85}$Rb atom is pumped. The spectral position of the pump frequency is indicated by a solid line; two thin vertical lines indicate the single scan.

Figures 4 (a),(b),(c) present fluorescence spectra taken in the cell when the $^{85}$Rb and $^{87}$Rb optical transitions were pumped.

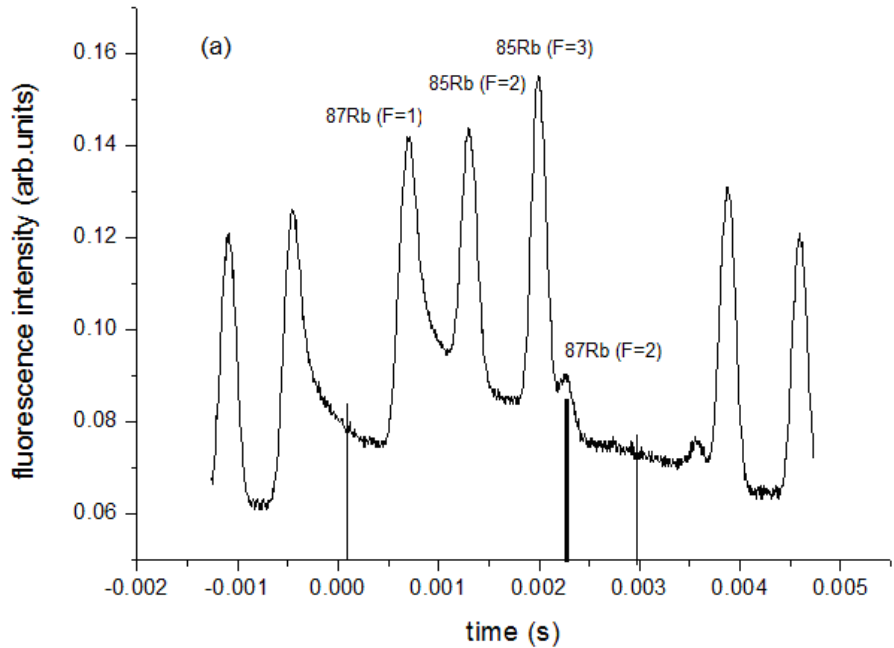

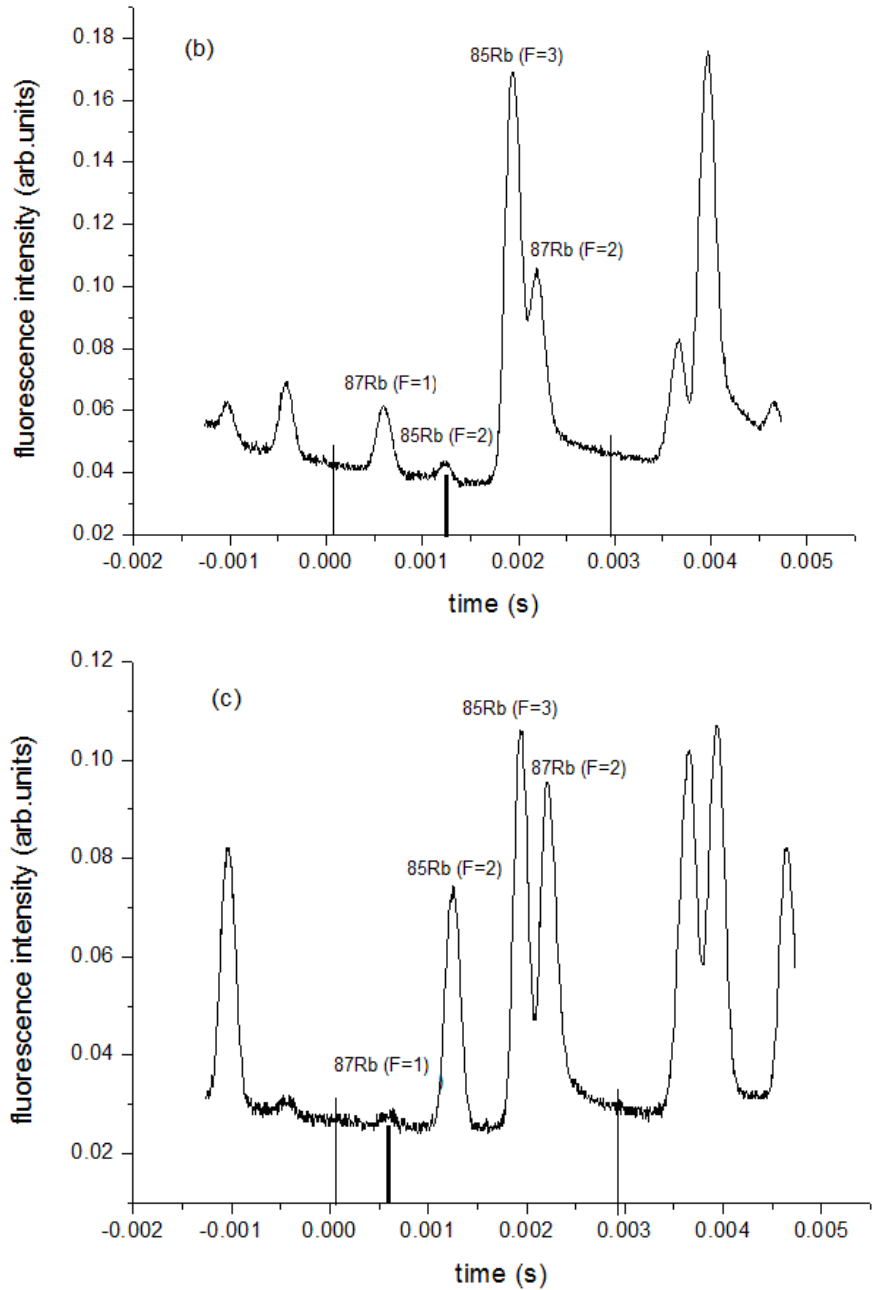

Figure 4. Fluorescence spectra recorded as a function of time when different optical transitions are pumped:
(a) - $^{87}$Rb $5S_{1/2}$ ($F = 2$) → $5P_{3/2}$,
(b) - $^{85}$Rb $5S_{1/2}$ ($F = 2$) → $5P_{3/2}$,
(c) - $^{87}$Rb $5S_{1/2}$ ($F = 1$) → $5P_{3/2}$,

 It is interesting to note that all spectra presented demonstrated similar spectral features with an equal decay time but with amplitude that depended strongly on the type of the pumping transition. This feature was large when the atoms were being "transported" by optical pumping from "upper" ground-state sublevel to a "lower" one: 5S1/2 (F = 3) → 5S1/2 (F = 2) of 85Rb or 5S1/2 (F = 2) → 5S1/2 (F = 1) of 87Rb; however it was relatively small when the atoms were pumped in the opposite direction - from the "lower" ground-state sublevel to the "upper" ground-state: 5S1/2 (F = 3) ← 5S1/2 (F = 2) of 85Rb or 5S1/2 (F = 2) ← 5S1/2 (F = 1) of 87Rb) (see, Figure 1 (b)). We must also admit that the isotope $^{87}$Rb displayed a somewhat larger effect than $^{85}$Rb.

We investigated these spectral features as a function of all important parameters, obtaining very similar results. In the following, only the measurements obtained by the pumping of $5S_{1/2}\,(F = 3) \rightarrow 5P_{3/2}$ optical transition $^{85}$Rb atom are discussed.

It was found that the shape of this specific fluorescence spectrum did not depend on the position of the detection point inside the cell or on the density of rubidium vapor that was controlled by changing the cell temperature from the operating temperature to zero $^0$C. This fact excluded the appearance of the observed spectral feature because of the re-absorption of the fluorescent light passing through the rubidium vapor. It was also found that the observed feature could be observed only in a high quality coated cell with a sufficiently long lifetime of optically pumped atoms at atomic sublevels in the ground state, as in our ceresin-coated cell with internal source of the atoms. For example, Figure 4 shows the same spectrum shown in Figure 3, but recorded in a paraffin-coated cell of the usual configuration. It can be clearly seen that, except for the deformed spectral peaks of the $^{85}$Rb atom, the recorded spectrum is practically devoid of this feature

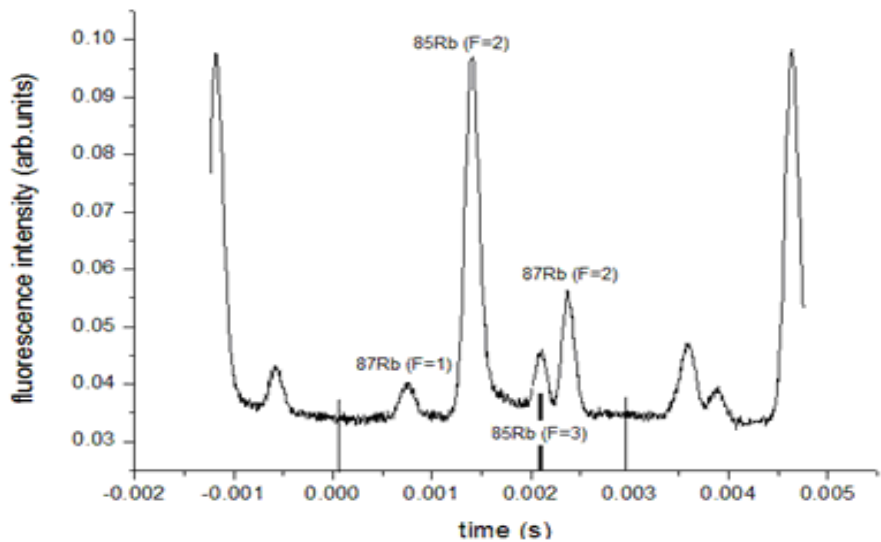

Figure 4. Fluorescence spectrum recorded as a function in paraffin-coated cell of usual configuration. It is recorded when the $5S_{1/2}\,(F = 3) \rightarrow 5P_{3/2}$ optical transition the $^{85}$Rb atom is pumped. The spectral position of the pump frequency is indicated by a solid line; two thin vertical lines indicate the single scan.

We changed the probing or pumping field intensities separately by putting a neutral density in these laser beams. Figures 5 (a), (b), (c) shows recorded spectra for probing laser powers of 30mW, 7,5 mW and 3mW respectively. It can be seen that with the decreasing of the intensity of the probing beam, only the population of the resonant level decreased, without any noticeable change in the spontaneous decay rate of the population of the resonant level.

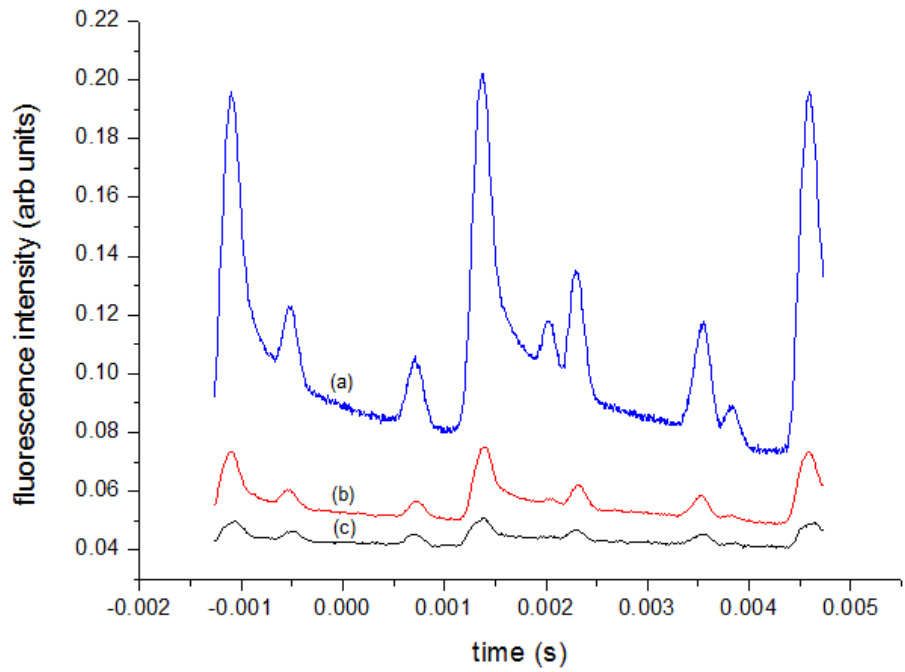

Figure 5. Fluorescence spectra recorded as a function of time
for different probing laser powers:
(a) - 30mW
(b) - 7,5 mW
(c) - 3mW

Figures 6 (a), (b), (c) show the recorded spectra for the pumping laser power of 30 mW, 7.5 mW and 3 mW respectively. It can be seen that an increase in the laser pumping power leads to an increase in the spontaneous decay time of the excited level.

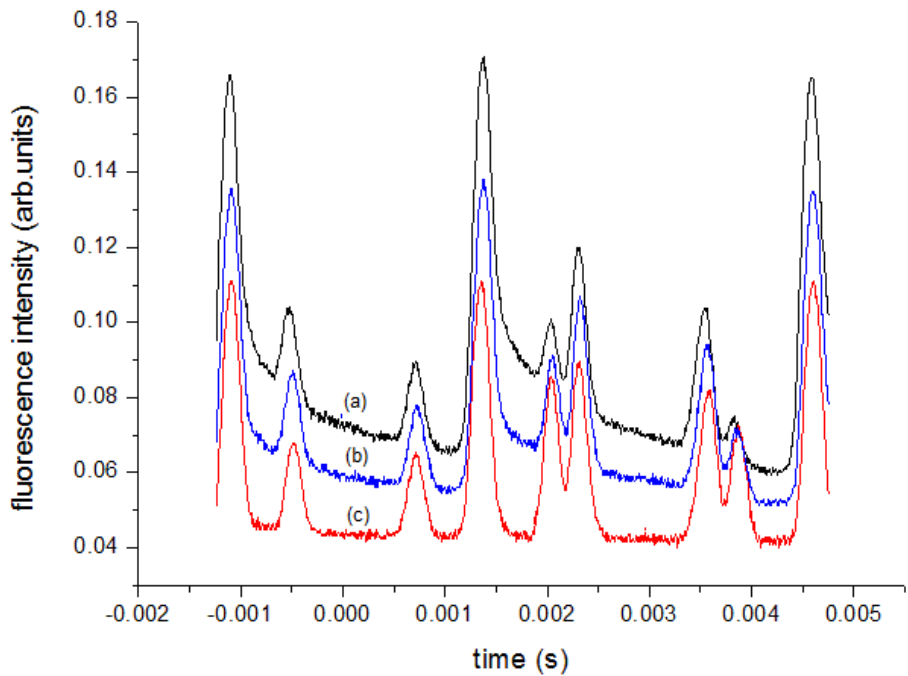

Figure 6. Fluorescence spectra recorded as a function of time
for different probing laser powers:
(a) - 30mW

(b) - 7,5 mW
(c) - 3mW

We now turn to a detailed discussion of the effects of the extension of the lifetime of the population on the atom excited state. The effect was modeled by solving rate equations that describe the dynamic problem of the interaction of intense bi-chromatic laser radiation with a 3-level Λ-scheme system for *F*, *mF* ground - and excited-state sublevels for Rb atom (see Fig.7).

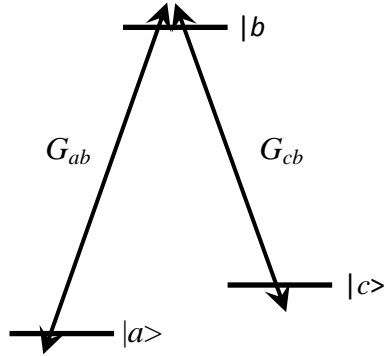

Figure. 7. 3-level model of the Rb atom

The levels | a> and | c> refer to the ground state sublevels that have a different energy as a result of the hyperfine splitting. The level | b> was excited by electro-dipole transitions from the ground sublevels | a> and | c>. Optical transitions between the levels | a> and | c> were strictly forbidden. To simplify matters, we considered the atoms to be immovable. The dynamics of the interaction of the two fields with an atom in an Λ-scheme is described by a standard system of equations for the density matrix. These rate equations include the optical pumping process by the pump beam, spontaneous emission; spin relaxation, and optical pumping by the probe beam, that can be written in the resonance approximation in the following form:

$$\frac{d}{dt}\hat{\rho}+\hat{\Gamma}\hat{\rho}=-\frac{i}{\hbar}[\hat{V}\hat{\rho}]+\hat{A}\hat{\rho}, \quad \hat{\rho}=\begin{pmatrix}\rho_{aa} & \rho_{ab} & \rho_{ac}\\ \rho_{ba} & \rho_{bb} & \rho_{bc}\\ \rho_{ca} & \rho_{cb} & \rho_{cc}\end{pmatrix}, \quad \hat{\Gamma}\hat{\rho}=\begin{pmatrix}\Gamma_{aa}\rho_{aa} & \Gamma_{ab}\rho_{ab} & \Gamma_{ac}\rho_{ac}\\ \Gamma_{ab}\rho_{ba} & \Gamma_{bb}\rho_{bb} & \Gamma_{cb}\rho_{bc}\\ \Gamma_{ac}\rho_{ca} & \Gamma_{cb}\rho_{cb} & \Gamma_{cc}\rho_{cc}\end{pmatrix},$$

(1)

$$\frac{\hat{V}}{\hbar}=\begin{pmatrix}0 & G_{ab}\exp(i\Omega_{ab}t) & 0\\ G_{ba}\exp(-i\Omega_{ab}t) & 0 & G_{bc}\exp(-i\Omega_{cb}t)\\ 0 & G_{cb}\exp(i\Omega_{cb}t) & 0\end{pmatrix}, \quad \hat{A}\hat{\rho}=\begin{pmatrix}A_{ba}\rho_{bb} & 0 & 0\\ 0 & 0 & 0\\ 0 & 0 & A_{bc}\rho_{bb}\end{pmatrix}.$$

where $\rho_{mn}=\rho_{nm}^{*}$ and $\Gamma_{mn}$ (m, n = a, b, c) are the matrix elements and their relaxation constants; Gab = Gba * and Gcb = Gbc * are the Rabi frequencies for the transitions a-b and c-b; $\Omega_{ab}$, $\Omega_{cb}$ - frequency detuning of laser fields from the frequencies of the corresponding optical transitions; $A_{ba}\rho_{bb}$ and $A_{bc}\rho_{bb}$ describe the increasing of the population on the ground state sublevels | a> and | c> due to the spontaneous decay from the | b> level.

We also assume that the frequency splitting of the ground state substantially exceeds the values of $\Omega_{ab}$ and $\Omega_{cb}$, i.e. each of the laser fields effectively interacts with one corresponding transition only. The application of the rotating field method leads to a system of differential equations for slow amplitudes:

$$\frac{d}{dt}\hat{r}+\hat{\Gamma}\hat{r}=-i[\hat{G}\hat{r}]+\hat{A}\hat{r},$$

$$\hat{r}=\begin{pmatrix} r_{aa} & r_{ab} & r_{ac} \\ r_{ba} & r_{bb} & r_{bc} \\ r_{ca} & r_{cb} & r_{cc} \end{pmatrix}=\begin{pmatrix} \rho_{aa} & \rho_{ab}\exp(i\Omega_{ab}t) & \rho_{ac}\exp[i(\Omega_{ab}-\Omega_{cb})t] \\ \rho_{ba}\exp(-i\Omega_{ab}t) & \rho_{bb} & \rho_{bc}\exp(-i\Omega_{cb}t) \\ \rho_{ca}\exp[i(\Omega_{cb}-\Omega_{ab})t] & \rho_{cb}\exp(i\Omega_{cb}t) & \rho_{cc} \end{pmatrix},$$

$$\hat{\Gamma}\hat{r}=\begin{pmatrix} \Gamma_{aa}r_{aa} & (\Gamma_{ab}+i\Omega_{ab})r_{ab} & (\Gamma_{ac}+i\Omega_{ab}-i\Omega_{cb})r_{ac} \\ (\Gamma_{ab}-i\Omega_{ab})r_{ba} & \Gamma_{bb}r_{bb} & (\Gamma_{cb}-i\Omega_{cb})r_{bc} \\ (\Gamma_{ac}-i\Omega_{ab}+i\Omega_{cb})r_{ca} & (\Gamma_{cb}+i\Omega_{cb})r_{cb} & \Gamma_{cc}r_{cc} \end{pmatrix}, \quad (2)$$

$$\hat{G}=\begin{pmatrix} 0 & G_{ab} & 0 \\ G_{ba} & 0 & G_{bc} \\ 0 & G_{cb} & 0 \end{pmatrix},\quad \hat{A}\hat{r}=\begin{pmatrix} A_{ba}r_{bb} & 0 & 0 \\ 0 & 0 & 0 \\ 0 & 0 & A_{bc}r_{bb} \end{pmatrix}.$$

The system of differential equations (2) has been solved numerically using the fourth-order Runge-Kutta method. At the initial time, the atom was in the ground state only. The remaining matrix elements were zero. We have divided the evolution into several stages. In the first stage, the laser fields were switched on and the Rabi Gab and Gcb frequencies were linearly increased from zero to the set values. In this case, the exact resonance condition held for both fields. In the second stage, both fields remained in exact resonance with the transitions a-b and c-b – that is, the frequencies of Rabi Gab and Gcb retained a constant set value. In the third stage, $\Omega_{cb}=0$ and $\Omega_{ab}$ was linearly increased from a zero value to a given large absolute positive value. In the fourth stage, the Gab laser field acting on the a-b transition was decreased from the pre-set value to zero. In the fifth stage, only the resonance field Gcb acted on the atom. All changes in the parameters are given by linear functions, and this has ensured the adiabatic character of the variation of these parameters and effect.

Figure 8 shows the dynamics of the population of the excited level in a three-level Λ-scheme for the Rabi frequency of the probing field Gab = 2, pumping field Gcb = 3. Note that this extremely simple model predicts the most general trends in the population dynamics remarkably well; it can be clearly seen that, after the change in the detuning Ωab, a rapid relaxation of ρbb was observed at time t ~ 13 arb.units, after which at time t > 13 arb.units the relaxation rate begins to slow down considerably. This is in qualitative accordance with the observation shown in Figure 3.

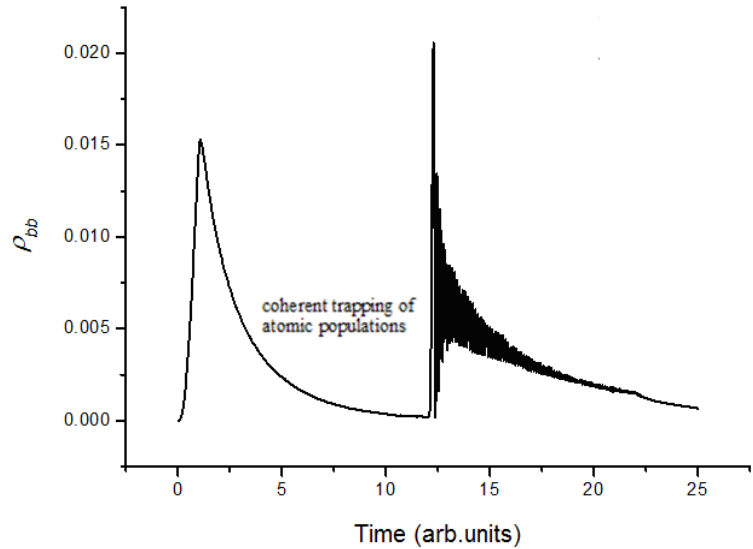

Figure.8. The dynamics of the population on the excited level:
The Rabi frequency of the probing field Gab = 2,
pumping field Gcb = 3.

Our model predicts that, as the intensity of the probing field decreases, only the population of the resonant level likewise decreases without changing their spontaneous decay rate. Figure 9 shows the dynamics of the population of the excited level in a three-level Λ-scheme with variations of the Rabi frequency of the probe field Gab.

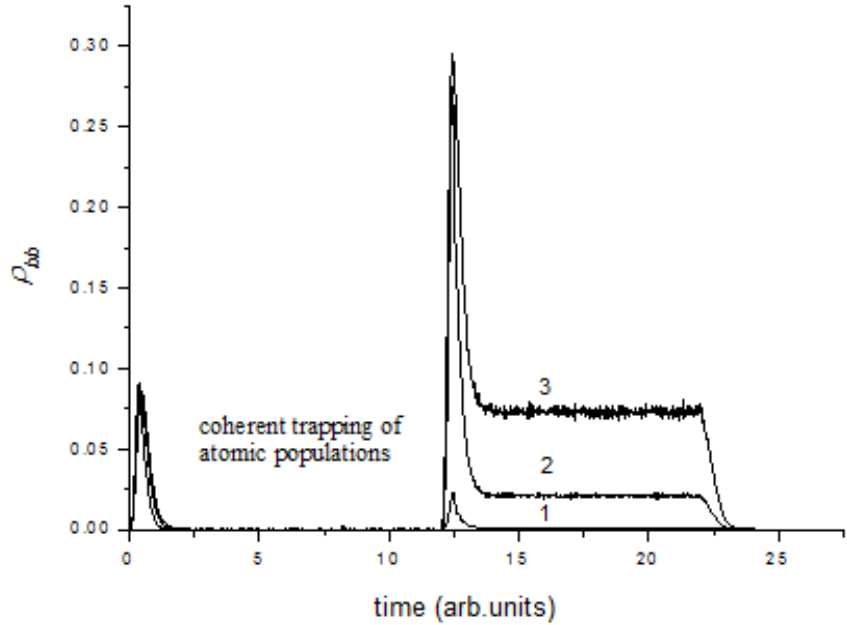

Figure.9. The dynamics of the population on the excited level:
1 - Gab = 2, 2 - Gab = 10, 3 - Gab = 20. The Rabi frequency
of the pumping field Gcb = 30.

It is clearly seen that, after the change in the detuning value $\Omega_{ab}$, a rapid relaxation of $\rho_{bb}$ is observed first, and then the relaxation rate becomes extremely small: at t> 13 arb.units, the dynamics diagrams of ρbb are practically "horizontal". It also can be seen that, as the intensity of the probing field decreases, only the population of the resonant level changes without any noticeable change in the spontaneous decay rate of the population of the resonant level. This is qualitatively consistent with our measurements shown in Figure 5.

Experiment and simulation agree well for low light pumping power with Gcb = 3. The model reproduces the narrow resonant peak together with the spectral long - decay feature and it shows that, with the decreasing of the intensity of the probing field, the observed feature decays without any noticeable change in its spontaneous decay rate. To obtain a nearly horizontal decay of the feature, it is necessary to use ten times higher pumping power up to about 300mW for Gcb = 30, as it is shown in Figure 9.

In conclusion, we have demonstrated surprising long - term spectral features in the fluorescence of $5S_{1/2} \rightarrow 5P_{3/2}$ the optical transition of both rubidium isotopes after we had adjusted the frequencies of both probing and pumping lasers. All experiments were made in low - density, room-temperature rubidium vapor. In a direct comparison with a paraffin-coated cell in the conventional configuration, we have demonstrated that the effect can only be observed in a high quality coated cell with a long lifetime of optical pumped atoms on atomic ground-state sublevels, for example, in our ceresin-coated cell with an internal atomic source. **Numerical simulation based on a complete density matrix is presented. It consistently demonstrates all the relevant features of the experiment. The model developed** supports the population pumping process through the ground-state sublevels of Rb atoms as the most plausible mechanism of the observed phenomenon. Future planned work will investigate this interesting effect in a cell coated with alkene compound that can support up to $10^6$ alkali-metal-wall collisions before depolarizing the alkali-metal spins [17].

The authors appreciate useful discussions with E. Podivilov. Special thanks for R. Robson (McKillop) for careful reading of the manuscript. This work was supported by the Research Program of Institute of Automation and Electrometry  SB RAS, number AAAA-A17-117060810014-9.